\documentclass[12pt,showpacs, showkeys]{revtex4}
\usepackage{graphicx}
\usepackage{amscd}
\usepackage{amsfonts}
\usepackage{amssymb}
\usepackage{amsmath}
\begin{document}
\baselineskip=18pt
\title{ Quark-antiquark and diquark condensates
in vacuum in two-flavor four-fermion interaction models with any
color number $N_c$ \footnote{The project supported by the National
Natural
Science Foundation of China under Grant No.10475113.} \\
}
\author{ZHOU Bang-Rong\footnote{E-mail:zhoubr@gucas.ac.cn}}
\affiliation{College of Physical Sciences, Graduate University of
the Chinese Academy of Sciences, Beijing 100049, China}

\date{}
\begin{abstract}
The color number $N_c$-dependence of the interplay between
quark-antiquark condensates $\langle\bar{q}q\rangle$ and diquark
condensates $\langle qq\rangle$ in vacuum in two-flavor four-fermion
interaction models is researched. The results show that the
$G_S$-$H_S$ (the coupling constant of scalar $(\bar{q}q)^2$-scalar
$(qq)^2$ channel) phase diagrams will be qualitatively consistent
with the case of $N_c=3$ as $N_c$ varies in 4D Nambu-Jona Lasinio
model and 2D Gross-Neveu (GN) model. However, in 3D GN model, the
behavior of the $G_S$-$H_P$ (the coupling constant of pseudoscalar
$(qq)^2$ channel) phase diagram will obviously depend on $N_c$. The
known characteristic that a 3D GN model has not the coexistence
phase of the condensates $\langle\bar{q}q\rangle$ and $\langle
qq\rangle$ is proven to appear only in the case of $N_c\leq 4$. In
all the models, the regions occupied by the phases containing the
diquark condensates $\langle qq\rangle$ in corresponding phase
diagrams will gradually decrease as $N_c$ grows up and finally go to
zero if $N_c \rightarrow \infty$, i.e. in this limit only the pure
$\langle\bar{q}q\rangle$ phase could exist.
\end{abstract}
\pacs{11.10.Lm, 11.10.Wx, 11.15.Pg, 11.30.Rd}
\keywords{four-fermion
interaction model, quark-antiquark and diquark condensates, color
number $N_c$ } \maketitle
\section{Introduction\label{Intro}}
The four-fermion interaction models including 4D Nambu-Jona-Lasinio
(NJL) model \cite{kn:1}, 3D and 2D Gross-Neveu (GN) model
\cite{kn:2} are not only ideal theoretical laboratories to research
dynamical symmetry breaking \cite{kn:1,kn:2,kn:3,kn:4,kn:5,kn:6} and
in recent years, the color-superconductivity \cite{kn:7,kn:8}, but
also can find many practical usefulness to Particle Physics and
Condensed Matter Physics. Hence it is certainly interesting to
conduct some more deepgoing exploration for these
field theory models themselves.\\
Let $q$ denote the fermion (called as quark) field and $\bar{q}$
antiquark field. Since in any four-fermion interaction model, the
couplings of $(qq)^2$-form and $(\bar{q}q)^2$-form can always
coexist via the Fierz transformation \cite{kn:8}, so once a
sufficiently strong coupling of $(qq)^2$-form leads to formation of
the diquark condensates $\langle qq\rangle$, then there will
certainly exist interplay between the quark-antiquark condensates
$\langle \bar{q}q\rangle$ and the diquark condensates $\langle
qq\rangle$ in the ground state of the system, even in vacuum
\cite{kn:9,kn:10,kn:11}. In vacuum condition, despite of absence of
net quarks, however, based on the relativistic quantum field theory,
it is possible that the diquark condensates $\langle qq\rangle$ and
di-antiquark condensates $\langle
\bar{q}\bar{q}\rangle$ are generated simultaneously.\\
\indent Motivated on the above idea, we have researched interplay
between the condensates $\langle \bar{q}q\rangle$ and $\langle
qq\rangle$ in vacuum in two-flavor 4D NJL \cite{kn:12}, 2D GN
\cite{kn:13} and 3D GN \cite{kn:14} models and find that the ratio
between the coupling constants of the $(\bar{q}q)^2$-form and the
$(qq)^2$-form plays a key role in determining the property of the
ground state. It should be indicated that some properties of the
system in the vacuum could be maintained to the case of finite
temperature and density. For instance, it has been shown that
\cite{kn:15}, in a 3D GN model with the quark's color number
$N_c=3$, the conclusion derived from the vacuum condition that the
model does not contain the phase with coexistence of the condensates
$\langle \bar{q}q\rangle$ and $\langle qq\rangle$ will be kept to
the case of finite temperature and finite quark chemical potential.
Therefore, the property of a model in vacuum could reflect its some
essential characteristic and this fact will make the research on
interplay between the two condensates in vacuum be given more
theoretical significance. In addition, a technical advantage is that
all the discussions can be made in an analytical way in this case.\\
\indent In our preceding work, the color number $N_c$ is fixed to be
3, consistent with phenomenology of Quantum Chromodynamics (QCD). In
this paper, we will extend the research to more general case of any
$N_c$. In fact, the large $N_c$ behavior of a four-fermion
interaction model is always an interesting problem to deserve to be followed.\\
\indent We will still use the effective potential approach employed
in Refs.\cite{kn:12,kn:13,kn:14} and obtain the expressions of
relevant effective potentials by similar procedures but omit the
details of these procedures. The mean field approximation will be
taken. In this case we will use  the Fierz transformed four-fermion
couplings in the Hartree approximation so as to avoid double
counting \cite{kn:8}. In selecting the couplings of the
$(qq)^2$-form, we will always simulate the $SU_c(N_c)$ gauge
interactions, where two fermions are attractive each other in the
antisymmetric $N_c(N_c-1)/2$-plet. The main results will also be
shown in some figures more clearly.
\section{4D Nambu-Jona-Lasinio model with any $N_c$\label{4DNJL}}
With 2 flavor and $N_c$ color massless fermions, the Lagrangian
\begin{eqnarray}
{\cal L}&=&\bar{q}i\gamma^{\mu}\partial_{\mu}q
+G_S[(\bar{q}q)^2+(\bar{q}i\gamma_5\vec{\tau}q)^2]\nonumber\\
&&+H_S\sum_{\lambda_A}(\bar{q}i\gamma_5\tau_2\lambda_Aq^C)
    ({\bar{q}}^Ci\gamma_5\tau_2\lambda_Aq),
\end{eqnarray}
where the fermion fields $q$ are in the doublet of $SU_f(2)$ and the
$N_c$-plet of $SU_c(N_c)$, i.e.
\begin{equation}
q=\left(
\begin{array}{c}
  u_i\\
  d_i \\
\end{array}
\right)\;\; i=1,\cdots,N_c,
\end{equation}
$q^C$  is the charge conjugate of $q$ and
$\vec{\tau}=(\tau_1,\tau_2,\tau_3)$ are the Pauli matrices acting in
two-flavor space. The matrices $\lambda_A$ run over all the
antisymmetric generators of $SU_c(N_c)$, $G_S$ and $H_S$ are
respectively the coupling constants in scalar $(\bar{q}q)^2$ channel
and scalar color
$N_c(N_c-1)/2$-plet $(qq)^2$ channel.\\
\indent Assume that the four-fermion interactions can lead to the
scalar condensates
\begin{equation}
\langle\bar{q}q\rangle  =\phi
\end{equation}
with all the $N_c$ color fermion entering them, and the scalar color
$N_c(N_c-1)/2$-plet difermion and di-antifermion condensates (after
a global $SU_c(N_c)$ transformation)
\begin{equation}
\langle{\bar{q}}^Ci\gamma_5\tau_2\lambda_2q\rangle=\delta,\;\;\;\;
\langle\bar{q}i\gamma_5\tau_2\lambda_2q^C\rangle=\delta^*,
\end{equation}
with only two color fermions  enter them. The corresponding symmetry
breaking is that $SU_{fL}(2)\otimes SU_{fR}(2)\rightarrow SU_f(2)$,
$SU_c(N_c)\rightarrow SU_c(2)$, and a "rotated" electric charge
$U_{\tilde{Q}}(1)$ and a "rotated" quark number $U'_q(1)$ leave
unbroken \cite{kn:8}. It should be indicated that in the case of
vacuum, the Goldstone bosons induced by spontaneous breaking of
$SU_c(N_c)$ could be some combinations of difermions and
di-antifermions \cite{kn:12}.\\
Define that
\begin{equation} \sigma=-2G_S\phi, \;\;
\Delta=-2H_S\delta,\;\;\Delta^*=-2H_S\delta^* .
\end{equation}
With  a 4D Euclidean momentum cutoff $\Lambda$, we can obtain the
relativistic effective potential \cite{kn:12}
\begin{widetext}\begin{eqnarray} V_4(\sigma,
|\Delta|)&=&\frac{\sigma^2}{4G_S}+\frac{|\Delta|^2}{4H_S}-\frac{1}{4\pi^2}
\left[(N_c\sigma^2+2|\Delta|^2)\Lambda^2-
(N_c-2)\dfrac{\sigma^4}{2}\left(\ln\frac{\Lambda^2}{\sigma^2}+\frac{1}{2}\right)\right.\nonumber\\
&&\left.-(\sigma^2+|\Delta|^2)^2\left(\ln\frac{\Lambda^2}{\sigma^2+|\Delta|^2}+\frac{1}{2}\right)\right],
\end{eqnarray}
\end{widetext}
where the conditions $\Lambda^2\gg \sigma^2$ and $\Lambda^2\gg
\sigma^2+\Delta^2$ have been assumed. Through an analytic
calculations, we have found that the ground states of the system,
i.e. the minimum points of $V_4(\sigma, |\Delta|)$, will be at
\begin{widetext}
\begin{equation} (\sigma,|\Delta|)=\left\{\begin{array}{lc}
  (0, & \Delta_1) \\
  (\sigma_2, & \Delta_2) \\
  (\sigma_1, & 0) \\
\end{array}\right.\;\;\mbox{if}\;\;
\left\{\begin{array}{lccl}
  &H_S\Lambda^2/\pi^2>1/2, \;\;\;\;\;\;\;\; 0&\leq&G_S/H_S<1/[1+(N_c-2)H_S\Lambda^2/\pi^2] \\
  &1/[1+(N_c-2)H_S\Lambda^2/\pi^2]&<&G_S/H_S<2/N_c \\
  &G_S\Lambda^2/\pi^2>1/N_c, &&G_S/H_S>2/N_c \\
\end{array}\right.,
\end{equation}
\end{widetext}
 Eq.(7) gives the phase diagram Fig.1 of the 4D NJL model.

\begin{figure}
{\includegraphics[width=0.5\textwidth,height=10cm]{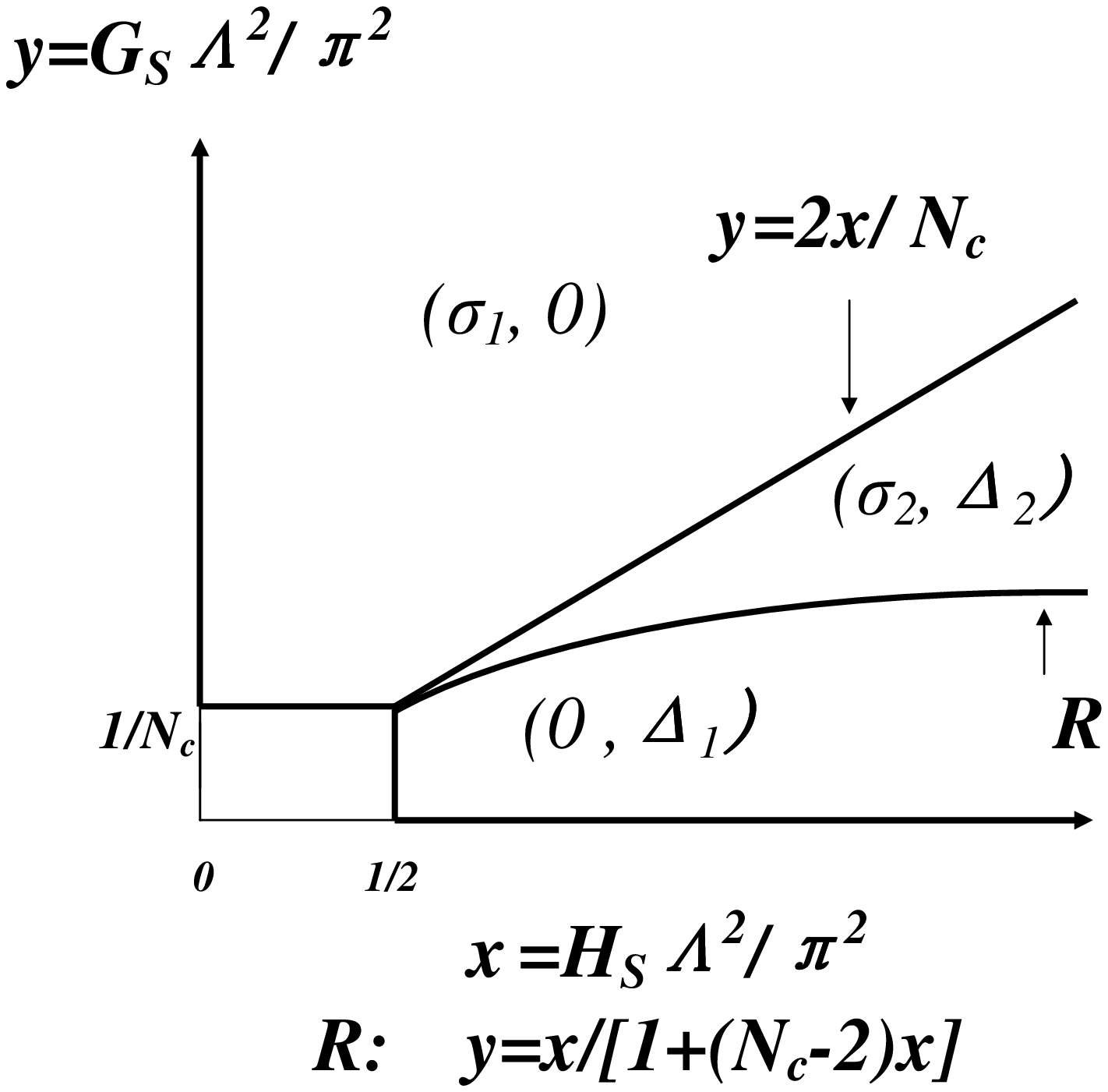}\\
\vspace{-2cm}\noindent {\footnotesize Fig.1: $G_S$-$H_S$ phase
diagram of a 4D Nambu-Jona-Lasinio model with the denotations that
$(0,\Delta_1)$--pure $\langle qq\rangle$ phase,
$(\sigma_2,\Delta_2)$--mixed phase with $\langle\bar{q}q\rangle$ and
     $\langle qq\rangle$ and $(\sigma_1,0)$--pure $\langle
     \bar{q}q\rangle$ phase.}}
\end{figure}

The results qualitatively coincides with the ones obtained in
Ref.\cite{kn:12} except that the color number is changed from 3 to
$N_c$. Now the pure $\langle \bar{q}q\rangle$ phase happens only if
the ratio $G_S/H_S>2/N_c$. This conclusion was obtained in
Ref.\cite{kn:5} but no $G_S-H_S$ phase diagram was given there. The
$2/N_c$ is just the ratio of the color numbers of the quarks
entering the diquark condensates and the quark-antiquark
condensates. The diquark condensates $\langle qq\rangle$ could
emergy only in the region where $H_S\Lambda^2/\pi^2>1/2$ and
$G_S/H_S<2/N_c$. It is seen from Fig.1 that as $G_S/H_S$ decreases
(or $H_S$ increases), we first come to a phase with coexistence of
the quark-antiquark condensates and the diquark condensates, then to
a pure diquark condensate phase, the boundary between the two phases
is determined by the curve $G_S=H_S/[1+(N_c-2)H_S\Lambda^2/\pi^2]$.
On the other hand, for a fixed $G_S$ and $H_S$, as $N_c$ grows up,
the slopes of the lines $G_S=2H_S/N_c$ and
$G_S=H_S/[1+(N_c-2)H_S\Lambda^2/\pi^2]$ will decrease. This implies
that the regions with the diquark condensates will become smaller
and smaller, and finally reduce to zeroes if $N_c\rightarrow
\infty$. In the latter limit, only pure quark-antiquark  condensates
could exist.
\section{2D Gross-Neveu model with any $N_c$\label{2DGN}}
The Lagrangian of the model is expressed by
\begin{eqnarray}
{\cal L}&=&\bar{q}i\gamma^{\mu}\partial_{\mu} q
+G_S[(\bar{q}q)^2+(\bar{q}i\gamma_5\vec{\tau}q)^2]\nonumber\\
&&+H_S(\bar{q}i\gamma_5\tau_S\lambda_Aq^C)
    ({\bar{q}}^Ci\gamma_5\tau_S\lambda_Aq),
\end{eqnarray}
All the denotations are the same as ones in 4D NJL model, except
that in 2D space-time $$ \gamma^0=\left(
\begin{array}{cc}
  1 & 0 \\
  0 & -1 \\
\end{array}
\right),\;\;\gamma^1=\left(
\begin{array}{cc}
  0 & 1 \\
  -1 & 0 \\
\end{array}
\right)=-C, \;\;\gamma_5=\gamma^0\gamma^1
$$
and $\tau_S=(\tau_0\equiv1, \tau_1, \tau_3)$ are flavor-triplet
symmetric matrices. It is indicated that the product matrix
$C\gamma_5\tau_S\lambda_A$ is antisymmetric.\\
\indent Assume that the four-fermion interactions could lead to the
scalar quark-antiquark condensates
\begin{equation}
\langle\bar{q}q\rangle=\phi,
\end{equation}
which will break the discrete symmetries
$$\chi_D:
q(t,x)\stackrel{\mathcal{\chi_D}}{\rightarrow}\gamma_5q(t,x),$$
$$\mathcal{P}_1:
q(t,x)\stackrel{\mathcal{P}_1}{\rightarrow}\gamma^1q(t,-x),$$ and
that the coupling with $H_S$ can lead to the scalar color
$N_c(N_c-1)/2$-plet difermion condensates and the scalar color anti-
$N_c(N_c-1)/2$-plet  di-antifermion condensates (after a global
transformation in flavor and color space)
\begin{equation}
\langle\bar{q}^Ci\gamma_51_f\lambda_2q\rangle=\delta,\;\;\;\;
\langle\bar{q}i\gamma_51_f\lambda_2q^C\rangle=\delta^*
\end{equation}
which will break discrete symmetries $Z_{N_c}^c$ (center of
$SU_c(N_c)$) down to $Z_2^c$, besides $\chi_D$ and
$\mathcal{P}_1$\cite{kn:13}. Noting that in a 2D model, no breaking
of continuous symmetry needs to be considered on the basis of
Mermin-Wagner-Coleman theorem\cite{kn:16}.\\
\indent The model is renormalizable. We may derive the renormalized
effective potential\cite{kn:13} containing the condensates (9) and
(10). In the space-time dimension regularization approach, we can
write down the renormalized $\mathcal{L}$ in $D=2-2\varepsilon$
dimension space-time by the replacements
$$
G_S\rightarrow G_SM^{2-D}Z_G,\; H_S\rightarrow H_SM^{2-D}Z_H,
$$
with the scale parameter $M$, the renormalization constants $Z_G$
and $Z_H$. In addition, the $\gamma^{\mu}$ in $\mathcal{L}$ will
become $2^{D/2}\times2^{D/2}$ matrices.\\
\indent Define the order parameters
\begin{equation}
\sigma=-2G_SM^{2-D}Z_G\phi, \;\; \;\; \Delta=-2H_SM^{2-D}Z_H\delta,
\end{equation}
which will be finite if $Z_G$ and $Z_H$ are selected so as to cancel
the UV divergences in $\phi$ and $\delta$. In the minimal
substraction scheme,
\begin{equation}
Z_G=1-\frac{2N_cG_S}{\pi}\frac{1}{\varepsilon},\;
Z_H=1-\frac{4H_S}{\pi}\frac{1}{\varepsilon}.
\end{equation}
The corresponding renormalized effective potential in the mean field
approximation up to one-loop order becomes
\begin{widetext}
\begin{eqnarray}
  V_2(\sigma, |\Delta|)&=& \frac{\sigma^2}{4G_S}
  -\frac{\sigma^2}{2\pi}\left(2\ln\frac{\bar{M}^2}{\sigma^2+|\Delta|^2}
  +(N_c-2)\ln\frac{\bar{M}^2}{\sigma^2}+N_c\right)\nonumber \\
  &&+\frac{|\Delta|^2}{4H_S}
  -\frac{|\Delta|^2}{\pi}\left(\ln\frac{\bar{M}^2}{\sigma^2+|\Delta|^2}+1\right),\;
  \bar{M}^2=2\pi e^{-\gamma}M^2,
\end{eqnarray}
\end{widetext}
where $\gamma$ is the Euler constant. $V_2(\sigma, |\Delta|)$
contains the two order parameters $\sigma$ and $|\Delta|$. By almost
identical demonstration to the one made in Ref.\cite{kn:13} (the
only change is that the color number of quark varies from 3 to
$N_c$), we obtain that the ground states of the system i.e. the
minimal points of $V_2(\sigma, |\Delta|)$ will be at
\begin{equation} (\sigma,|\Delta|)=\left\{\begin{array}{lc}
  (0, & \Delta_1) \\
  (\sigma_2, & \Delta_2) \\
  (\sigma_1, & 0) \\
\end{array}\right.\;\;\mbox{if}\;\;
\left\{\begin{array}{c}
  G_S/H_S=0 \\
  0<G_S/H_S<2/N_c \\
  G_S/H_S>2/N_c \\
\end{array}\right.
\end{equation}
The results (14) can be drawn in Fig.2 which is the $G_S-H_S$ phase
diagram of the 2D GN model.\\

\begin{figure}
     {\centering
     \includegraphics[width=0.5\textwidth, height=10.5cm]{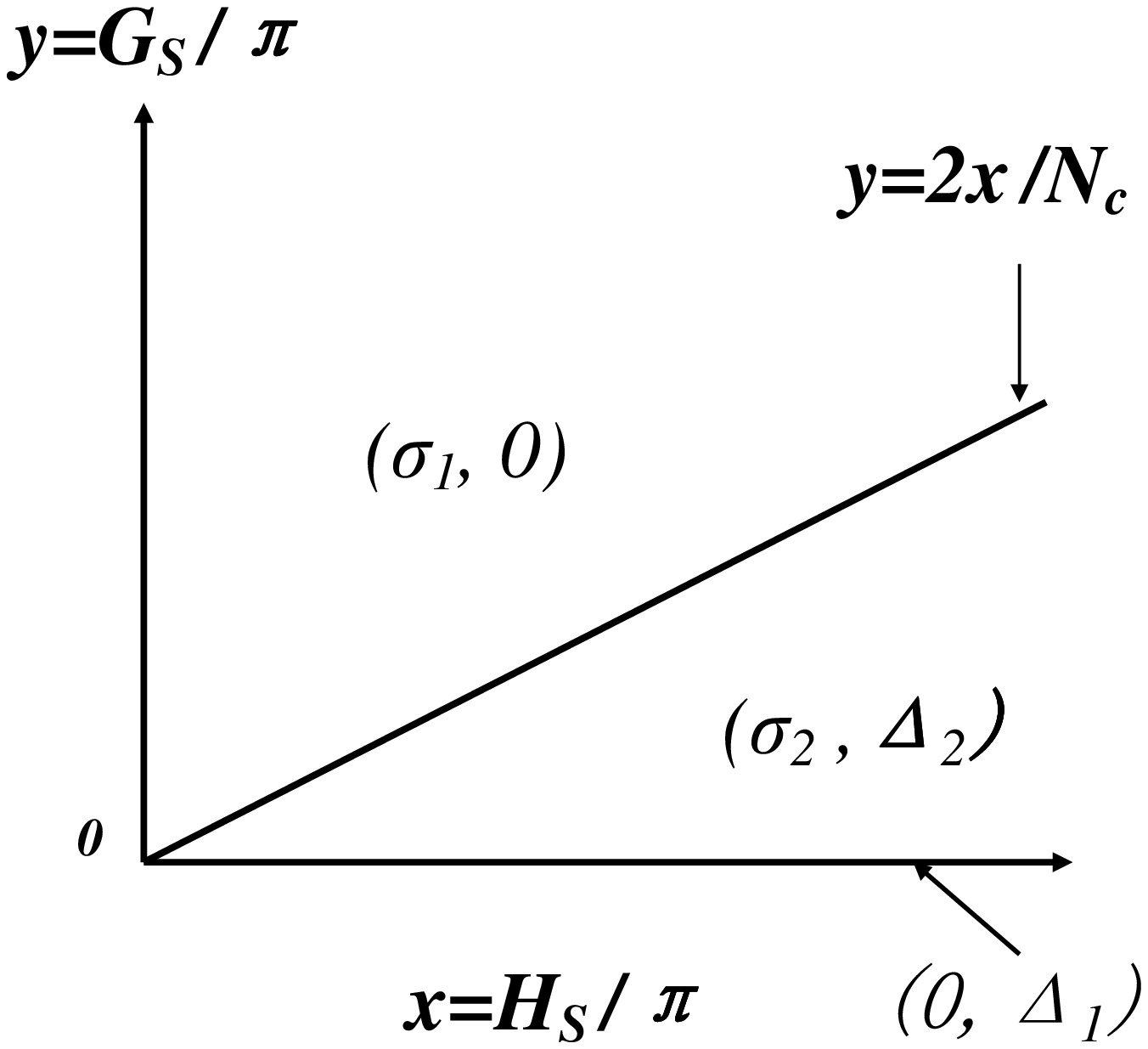}\\
     \vspace{-3.5cm}\noindent{\footnotesize Fig.2: $G_S-H_S$ phase diagram of a 2D Gross-Neveu model with the
     denotations that $(0,\Delta_1)\;-$ pure $\langle
     qq\rangle$ phase, $(\sigma_2,\Delta_2)\;-$ mixed phase with $\langle\bar{q}q\rangle$ and
     $\langle qq\rangle$ and $(\sigma_1,0)\;-$ pure $\langle
     \bar{q}q\rangle$ phase.}}
\end{figure}

 The phase diagram are similar to the one obtained in
the case of $N_c=3$, including that formations of the condensates do
not call for the coupling constant $G_S$ and $H_S$ having some lower
bounds, however, the boundary between the pure quark-antiquark
condensate phase and the coexistence phase with quark-antiquark
condensate and diquark condensates is now replaced by
$G_S/H_S=2/N_c$. It is seen that as $N_c$ increases, the area in
Fig.2 occupied by the coexistence phase of quark-antiquark
condensate and diquark condensates will gradually decrease and
finally disappear if $N_c\rightarrow \infty$.  It is indicated that
if $N_c\rightarrow\infty$, the region corresponding to the pure
diquark condensates in the diagram (the transverse axis of Fig.2)
will also not exist. In fact, the pure diquark condensate phase
corresponds to the minimal point of $V_2(\sigma,|\Delta|)$
$(\sigma,|\Delta|)=(0,\Delta_1)$, where $\Delta_1$ is the non-zero
solution of the equation $\partial
V_2(\sigma,|\Delta|)/\partial|\Delta|=0$ with $\sigma=0$. In this
case, it is easy to find that the determinant of the second
derivatives of $V_2$
\begin{equation*}
    K_2=\left|\begin{array}{cc}
                  A_2 & B_2 \\
                  B_2 & C_2 \\
                \end{array}
                  \right|
\end{equation*}
with $A_2=\partial^2V_2/\partial\sigma^2$,
$B_2=\partial^2V_2/\partial\sigma/\partial|\Delta|=
\partial^2V_2/\partial|\Delta|/\partial\sigma$ and
$C_2=\partial^2V_2/\partial|\Delta|^2$. For the solution
$(0,\Delta_1)$, we get that
\begin{equation*}
A_2=\dfrac{1}{2G_S}-\frac{1}{2H_S}-
\left.\dfrac{N_c-2}{\pi}\ln\dfrac{\bar{M}^2}{\sigma^2}\right|_{\sigma\rightarrow
0}+\dfrac{2(N_c-2)}{\pi},
\end{equation*}
\begin{equation*}
    K_2=4A_2/\pi.
\end{equation*}
If $N_c$ is finite, then $G_S=0$ could make $A_2>0$ and $K_2>0$ thus
the solution $(0,\Delta_1)$ becomes a minimal point of $V_2$.
However, if $N_c\rightarrow\infty$, then the third term of $A_2$
will have the $-\infty^2$ behavior, hence even if $G_S=0$ then it is
impossible to have $A_2>0$. As a result, the solution $(0,\Delta_1)$
is no longer a extreme point of $V_2$. This will imply that the
region of the pure diquark condensate phase corresponding to the
transverse axis of Fig.2 will not exist for finite $H_S$, $G_S=0$
and $N_c\rightarrow\infty$.
 \vspace{1cm}
\section{3D Gross-Neveu
model with any $N_c$}\label{3dGN} The Lagrangian of the model is
 expressed by
\begin{eqnarray}
{\cal L}&=&\bar{q}i\gamma^{\mu}\partial_{\mu} q
+G_S[(\bar{q}q)^2+(\bar{q}\vec{\tau}q)^2]\nonumber\\
&&+H_P\sum_{\lambda_A}(\bar{q}\tau_2\lambda_Aq^C)
    ({\bar{q}}^C\tau_2\lambda_Aq),
\end{eqnarray}
where $\gamma^{\mu} (\mu=0,1,2)$ are taken to be $2\times 2$
matrices
$$ \gamma^0=\left(
\begin{array}{cc}
  1 & 0 \\
  0 & -1 \\
\end{array}
\right),\;\gamma^1=\left(
\begin{array}{cc}
  0 & i \\
  i & 0 \\
\end{array}
\right),\;\gamma^2=\left(
\begin{array}{cc}
  0 & 1 \\
  -1 & 0 \\
\end{array}
\right)=C.
$$
It is noted that the product matrix $C\tau_2\lambda_A$ is
antisymmetric, and since without the $"\gamma_5"$ matrix, the only
possible color $N_c(N_c-1)/2$-plet difermion interaction channel is
pseudoscalar one. The condensates $\langle \bar{q}q\rangle$ will
break time reversal symmetry $\mathcal{T}: q(t,\vec{x})\rightarrow
\gamma^2q(-t,\vec{x})$, special parity $\mathcal{P}_1:
q(t,x^1,x^2)\rightarrow \gamma^1 q(t,-x^1,x^2)$, special parity $
\mathcal{P}_2: q(t,x^1,x^2)\rightarrow \gamma^2 q(t,x^1,-x^2)$. The
difermion condensates $\langle{\bar{q}}^C\tau_2\lambda_2q\rangle$
(after a global rotation in the color space) will break $
SU_c(N_c)\rightarrow SU_c(2) $ and leave a "rotated" electrical
charge $U_{\tilde{Q}}(1)$ and a "rotated" fermion number $U'_q(1)$
unbroken. It also breaks $ \mathrm{parity}\;\; \mathcal{P}:
q(t,\vec{x})\rightarrow \gamma^0 q(t,-\vec{x}) $
and this shows pseudoscalar feature of the difermion condensates.\\
\indent Define the order parameters in the 3D GN model
\begin{equation}
\sigma = -2G_S\langle\bar{q}q\rangle, \; \; \;
 \Delta = -2H_P\langle\bar{q}^C\tau_2\lambda_2q\rangle,
\end{equation}
 The effective potential in the mean field
approximation can be expressed by \cite{kn:14}
\begin{eqnarray}
 V_3(\sigma, |\Delta|)&=& \frac{\sigma^2}{4G_S}+\frac{|\Delta|^2}{4H_P}
 -\frac{1}{\pi^2}(N_c\sigma^2+2|\Delta|^2)\Lambda_3\nonumber \\
 &&+\frac{1}{3\pi}\left[6\sigma^2|\Delta|+2|\Delta|^3+(N_c-2)\sigma^3\right.\nonumber \\
 &&\left.+ 2\theta(\sigma-|\Delta|)(\sigma-|\Delta|)^3\right],
\end{eqnarray}
where $\Lambda_3$ is a 3D Euclidean momentum cutoff and the
assumptions $\Lambda_3\gg |\sigma-|\Delta||$, $\Lambda_3\gg
\sigma+|\Delta|$ and $\Lambda_3\gg \sigma$ have been made. The
ground states of the system correspond to the least value points of
$V_3(\sigma, |\Delta|)$. Since the phase structure of the 3D GN
model will have qualitatively different $N_c$-dependence from the
one of the 4D and 2D models, we will make more detailed discussions
of this problem.\\
\indent The extreme value conditions $\partial V_3(\sigma,
|\Delta|)/\partial \sigma=0$ and $\partial V_3(\sigma,
|\Delta|)/\partial |\Delta|=0$ will respectively become
\begin{eqnarray}
&&\sigma\left(\frac{1}{2G_S}-\frac{2N_c\Lambda_3}{\pi^2}
+\frac{4|\Delta|}{\pi}+\frac{(N_c-2)\sigma}{\pi}\right)\nonumber\\
&&+\frac{2}{\pi}\theta(\sigma-|\Delta|)(\sigma-|\Delta|)^2=0,\vspace{0.5cm}
\end{eqnarray}
\begin{eqnarray}
&&|\Delta|\left(\frac{1}{2H_P}-\frac{4\Lambda_3}{\pi^2}+
\frac{2|\Delta|}{\pi}\right)\nonumber\\
&&+\frac{2}{\pi}[\sigma^2-\theta(\sigma-|\Delta|)(\sigma-|\Delta|)^2]=0.
\end{eqnarray}
Define the determinant from the second derivatives of $V_3(\sigma,
|\Delta|)$
$$
K=\left|\begin{array}{cc}
  A & B \\
  B & C \\
\end{array}\right|=AC-B^2,
$$
where
\begin{widetext}
\begin{eqnarray}
A&=&\frac{\partial^2V_3}{\partial\sigma^2}=\frac{1}{2G_S}-\frac{2N_c\Lambda_3}{\pi^2}
+\frac{4|\Delta|}{\pi}+\frac{2(N_c-2)\sigma}{\pi}
+\frac{4}{\pi}\theta(\sigma-|\Delta|)(\sigma-|\Delta|),
  \nonumber \\
B&=&\frac{\partial^2V_3}{\partial\sigma\partial|\Delta|}=
\frac{\partial^2V_3}{\partial|\Delta|\partial\sigma}=
\frac{4}{\pi}[\sigma-\theta(\sigma-|\Delta|)(\sigma-|\Delta|)],\nonumber \\
C&=&\frac{\partial^2V_3}{\partial|\Delta|^2}=
\frac{1}{2H_P}-\frac{4\Lambda_3}{\pi^2} +\frac{4|\Delta|}{\pi}
+\frac{4}{\pi}\theta(\sigma-|\Delta|)(\sigma-|\Delta|).
\end{eqnarray}
\end{widetext}
The equations (18) and (19) have the following four different
solutions which will be discussed in order. \\
\indent (i) ($\sigma, |\Delta|$)=(0,0). It is a maximum point of
$V_3(\sigma, |\Delta|)$, since in this case we have
\begin{eqnarray*}
A&=&1/2G_S-2N_c\Lambda_3/\pi^2<0 ,\\
K&=&A\left(1/2H_P-4\Lambda_3/\pi^2\right)>0,
\end{eqnarray*}
assuming Eqs. (18) and (19) have solutions of non-zero $\sigma$ and
$|\Delta|$. \\
\indent (ii) ($\sigma, |\Delta|$)=($\sigma_1$,0), where the non-zero
$\sigma_1$ satisfies the equation
\begin{equation}
1/2G_S-2N_c\Lambda_3/\pi^2+N_c\sigma_1/\pi=0,
\end{equation}
under the condition that $G_S\Lambda_3/\pi^2>1/4N_c$. For this
solution, by means of Eq.(21) we obtain
\begin{eqnarray*}
  A &=&N_c\sigma_1/\pi, \\
  K&=&A\left(\frac{1}{2H_P}-\frac{1}{N_cG_S}+\frac{2\sigma_1}{\pi}\right)>0,
  \;\; \mathrm{if}\;\; \frac{G_S}{H_P}>\frac{2}{N_c}.
\end{eqnarray*}
Hence ($\sigma_1$,0) is a minimum point of $V_3(\sigma,
|\Delta|)$ when $G_S/H_P>2/N_c$.\\
\indent (iii) ($\sigma$, $|\Delta|$)= (0, $\Delta_1$), where
non-zero $\Delta_1$ obeys the equation
\begin{equation}
1/2H_P-4\Lambda_3/\pi^2+2\Delta_1/\pi=0,
\end{equation}
under the condition that $H_P\Lambda_3/\pi^2>1/8$. For this
solution, by using Eq.(22), we get
\begin{eqnarray}
 A&=&\frac{1}{2G_S}-\frac{N_c}{4H_P}+(4-N_c)\frac{\Delta_1}{\pi}\\
   &=& \frac{1}{2G_S}-\frac{2N_c\Lambda_3}{\pi^2}+\frac{4\Delta_1}{\pi},\\
K&=&2\Delta_1A/\pi.
\end{eqnarray}
\indent (a) When $N_c\leq 4$, from Eqs.(23) and (25) we have
\begin{equation*}
A>0,K>0 \;\;\mathrm{for} \;\;\frac{G_S}{H_p}<\frac{2}{N_c},
\frac{H_P\Lambda_3}{\pi^2}>\frac{1}{8}.
\end{equation*}
\indent (b) When $N_c> 4$, from Eqs.(24) and (25) we have
\begin{equation*}
A>0,K>0 \;\;\mathrm{for}\;\;
\frac{G_S\Lambda_3}{\pi^2}<\frac{1}{4N_c},
\frac{H_P\Lambda_3}{\pi^2}>\frac{1}{8}.
\end{equation*}
Noting that in this case the constraint $G_S/H_P<2/N_c$ is also
implied owing to that
\begin{equation*}
1/8H_P<\Lambda_3/\pi^2<1/4N_cG_S.
\end{equation*}
Thus in the cases of both (a) and (b), the solution
$(\sigma,|\Delta|)=(0,\Delta_1)$ is a minimum value point of
$V_3(\sigma,|\Delta|)$. \\
\indent (iv) $(\sigma,|\Delta|)=(\sigma_2,\Delta_2)$. In view of the
function $\theta(\sigma-|\Delta|)$ in Eqs.(18) and (20), we have to
consider the two cases of $\sigma_2>\Delta_2$ and
$\sigma_2<\Delta_2$
respectively. \\
\indent (a) $\sigma_2>\Delta_2$. In this case Eqs.(18) and (19) will
lead to
\begin{equation*}
\frac{1}{2G_S}-\frac{N_c}{4H_P}=\frac{N_c\sigma_2^2-2\Delta_2^2}{\pi\sigma_2}>0\;\;
\mathrm{thus}\;\;\frac{G_S}{H_P}<\frac{2}{N_c},
\end{equation*}
since $N_c\geq2$. Then we obtain from Eq.(20) that
\begin{equation*}
A=\frac{N_c\sigma_2^2-2\Delta_2^2}{\pi\sigma_2}>0
\;\;\mathrm{and}\;\;K=-\frac{16\Delta_2^2}{\pi^2}<0.
\end{equation*}
Hence the solution $(\sigma_2,\Delta_2)$ is not an extreme value
point of $V_3(\sigma,|\Delta|)$ if $\sigma_2>\Delta_2$.\\
\indent (b) $\sigma_2<\Delta_2$. In this case Eqs.(18) and (19)
become
\begin{equation}
\frac{1}{2G_S}-\frac{2N_c\Lambda_3}{\pi^2}
+\frac{4\Delta_2+(N_c-2)\sigma_2}{\pi}=0,
\end{equation}
and
\begin{equation} \frac{1}{2H_P}-\frac{4\Lambda_3}{\pi^2}+
\frac{2\Delta_2}{\pi}+\frac{2\sigma_2^2}{\pi\Delta_2}=0
\end{equation}
which may lead to the equations
\begin{widetext}
\begin{eqnarray}
 \frac{1}{2H_P}-\frac{1}{N_cG_S}&=&\frac{2}{\pi N_c\Delta_2}
 \left[(4-N_c)\Delta_2^2+(N_c-2)\sigma_2\Delta_2-N_c\sigma_2^2\right] \nonumber\\
  &=&\frac{2(4-N_c)}{\pi N_c\Delta_2}\left[\left(\Delta_2+\frac{N_c-2}{2(4-N_c)}\sigma_2\right)^2
  +\frac{4N_c(N_c-4)-(N_c-2)^2}{4(4-N_c)^2}\sigma_2^2\right]
  \end{eqnarray}
\end{widetext}
and Eq.(20) will lead to that
\begin{eqnarray}
  K&=&A\frac{2}{\pi\Delta_2}\left(\Delta_2^2-\sigma_2^2-\frac{8}{N_c-2}\Delta_2\sigma_2\right) \nonumber \\
   &=& A\frac{2}{\pi\Delta_2}\left(\Delta_2+\frac{[16+(N_c-2)^2]^{\frac{1}{2}}-4}{N_c-2}\sigma_2\right)\tilde{K} \nonumber \\
  A&=&(N_c-2)\frac{\sigma_2}{\pi}  \nonumber \\
  \tilde{K}&=& \Delta_2-\frac{[16+(N_c-2)^2]^{\frac{1}{2}}+4}{N_c-2}\sigma_2
\end{eqnarray}
Obviously, for $N_c>2$, $\tilde{K}>0$ will be the condition in which
the solution $(\sigma_2, \Delta_2)$ becomes a minimum value point of
$V_3(\sigma,|\Delta|)$. By means of Eqs.(28) and (29) we may discuss
the relation between $G_S$ and $H_P$ when the $(\sigma_2,\Delta_2)$
becomes a possible minimum value point of $V_3(\sigma,|\Delta|)$.
The result is obviously $N_c$-dependent.\\
\indent (b1) $N_c=2$. In this case $A=0$, $K=-16\sigma_2^2/\pi^2<0$,
so $(\sigma_2,\Delta_2)$ is not an extreme point of $V_3$.\\
\indent (b2) $N_c=3$. In this case, $\tilde{K}>0$ or $K>0$ become
$\Delta_2>(\sqrt{17}+4)\sigma_2$ or
$\Delta_2^2>\sigma_2^2+8\sigma_2\Delta_2$. Substituting these
inequalities into Eq.(28) we get
\begin{eqnarray*}
 \frac{1}{2H_P}-\frac{1}{N_cG_S}&=&\frac{2}{3\pi\Delta_2}(\Delta_2^2+\sigma_2\Delta_2-3\sigma_2^2)  \\
   &>& \frac{3\sigma_2^2}{\pi\Delta_2}(\sqrt{17}+4-\frac{2}{9})>0. \\
\end{eqnarray*}
\indent (b3) $N_c=4$. In this case, $\tilde{K}>0$ becomes
$\Delta_2>(\sqrt{5}+2)\sigma_2$, then we have
\begin{equation*}
\frac{1}{2H_P}-\frac{1}{N_cG_S}=\frac{\sigma_2}{\pi\Delta_2}(\Delta_2-2\sigma_2)>
\frac{\sqrt{5}\sigma_2^2}{\pi\Delta_2}>0.
\end{equation*}
The results in (b2) and (b3) indicate that when $N_c=3,4$ and
$G_S/H_P>2/N_c$, the non-zero solution $(\sigma_2,\Delta_2)$ of
Eqs.(26) and (27) could be a minimum point of
$V_3(\sigma,|\Delta|)$.\\
\indent we have obtained in (ii) that when $G_S/H_P>2/N_c$, the
non-zero solution $(\sigma_1,0)$ of Eqs.(18) and (19) is also a
minimum point of $V_3(\sigma,|\Delta|)$ and the result is
$N_c$-independent. In this case we must determine which one of the
solutions $(\sigma_2,\Delta_2)$ and $(\sigma_1,0)$ is the least
value point of $V_3(\sigma,|\Delta|)$. For this purpose, we will
compare the value of $V_3(\sigma,|\Delta|)$ at the point
$(\sigma_1,0)$ with the one at the point $(\sigma_2,\Delta_2)$. In
fact, by using Eqs.(17) and (21), we may obtain that
\begin{equation}
V_3(\sigma_1,0)=-N_c\sigma_1^3/6\pi
\end{equation}
and by using Eqs.(17),(26)and (27), we can get  that
\begin{equation}
\left.V_3(\sigma_2,\Delta_2)\right|_{\sigma_2<\Delta_2}=
-\frac{1}{3\pi}\left(\Delta_2^3+3\Delta_2\sigma_2^2+\frac{N_c-2}{2}\sigma_2^3\right).
\end{equation}
In addition, comparing Eq.(21) with Eq.(26) with $G_S$ and $H_P$
being fixed, we will be led to the relation that
\begin{equation}
N_c\sigma_1=4\Delta_2+(N_c-2)\sigma_2
\end{equation}
By means of Eq.(32) we may calculate that
\begin{widetext}
\begin{eqnarray}
  V_3(\sigma_1,0)-V_3(\sigma_2,\Delta_2)
  &=&-\frac{1}{3\pi}\left\{
  \begin{array}{ll}
    \left[\frac{1}{9}(23\Delta_2^3-4\sigma_2^3)+\frac{1}{3}\sigma_2\Delta_2(8\Delta_2-7\sigma_2)\right],\;\;\;\;
     \mathrm{if}\;\;N_c=3 \vspace{0.3cm}\\
    \left[\Delta_2^3-\frac{3}{4}\sigma_2^3+\frac{1}{2}\sigma_2\Delta_2(2\Delta_2-\sigma_2)\right],\;\;\;\;  \mathrm{if}\;\;N_c=4 \\
  \end{array}
  \right. \nonumber \\
  &<&0\;\;\;\;\mathrm{for}\;\;\sigma_2<\Delta_2
\end{eqnarray}
\end{widetext}
This shows that when $N_c=3,4$ and $G_S/H_P>2/N_c$, the least value
point of $V_3(\sigma,|\Delta|)$ is $(\sigma_1,0)$ rather than
$(\sigma_2,\Delta_2)$.  The same conclusion is also true for the
case of $N_c=2$ owing to the fact that $(\sigma_2,\Delta_2)$ is now
not an extreme value point of $V_3(\sigma, |\Delta|)$, as shown in
(b1). These results indicate that, when $G_S/H_P>2/N_c$, only the
pure quark-antiquark condensates could exist in the ground state of
the
3D GN model. \\
\indent (b4) $N_c>4$. In this case, the solution
$(\sigma_2,\Delta_2)$ which satisfies $\tilde{K}>0$ will be a
minimum value point of $V_3(\sigma, |\Delta|)$. Noting that the
right-handed side of Eq.(28) is always less than zero when  $N_c>4$
and this implies that
\begin{equation*}
G_S/H_P<2/N_c \;\;\mathrm{when}\;\;N_c>4,
\end{equation*}
hence we may affirm that if the solution $(\sigma_2,\Delta_2)$ is a
minimum value point of $V_3(\sigma, |\Delta|)$ when $N_c>4$, then
the condition $G_S/H_P<2/N_c$ must be obeyed.\\
\indent Now the least value points of $V_3(\sigma, |\Delta|)$ can be
summed up as follows:
\begin{widetext}
\begin{equation}
(\sigma, |\Delta|)= \left\{ \begin{array}{c}
  (0, \Delta_1), \\
   \left\{\begin{array}{c}
     (0, \Delta_1),\\
     (\sigma_2,\Delta_2), \\
   \end{array}
   \right. \\
  (\sigma_1,0), \\
\end{array}\right.
\begin{array}{lllll}
  \mathrm{if} & G_S/H_P<2/N_c, &H_P\Lambda_3/\pi^2>1/8, &  & \mathrm{for}\; N_c\leq 4\\
  \mathrm{if} & G_S/H_P<2/N_c, & H_P\Lambda_3/\pi^2>1/8,& G_S\Lambda_3/\pi^2
\left\{\begin{array}{l}
  <1/4N_c, \\
  >1/4N_c, \\
\end{array}
  \right.  & \mathrm{for}\; N_c>4 \\
  \mathrm{if} & G_S/H_P>2/N_c,  & &G_S\Lambda_3/\pi^2>
1/4N_c, & \mathrm{for\; all}\; N_c \\
\end{array}
\end{equation}
\end{widetext}
Eq.(34) gives the $G_S-H_P$ phase diagrams Fig.3 and Fig.4 of the 3D
GN model.

\begin{figure}
     {\centering
     \includegraphics[width=0.5\textwidth,height=11cm]{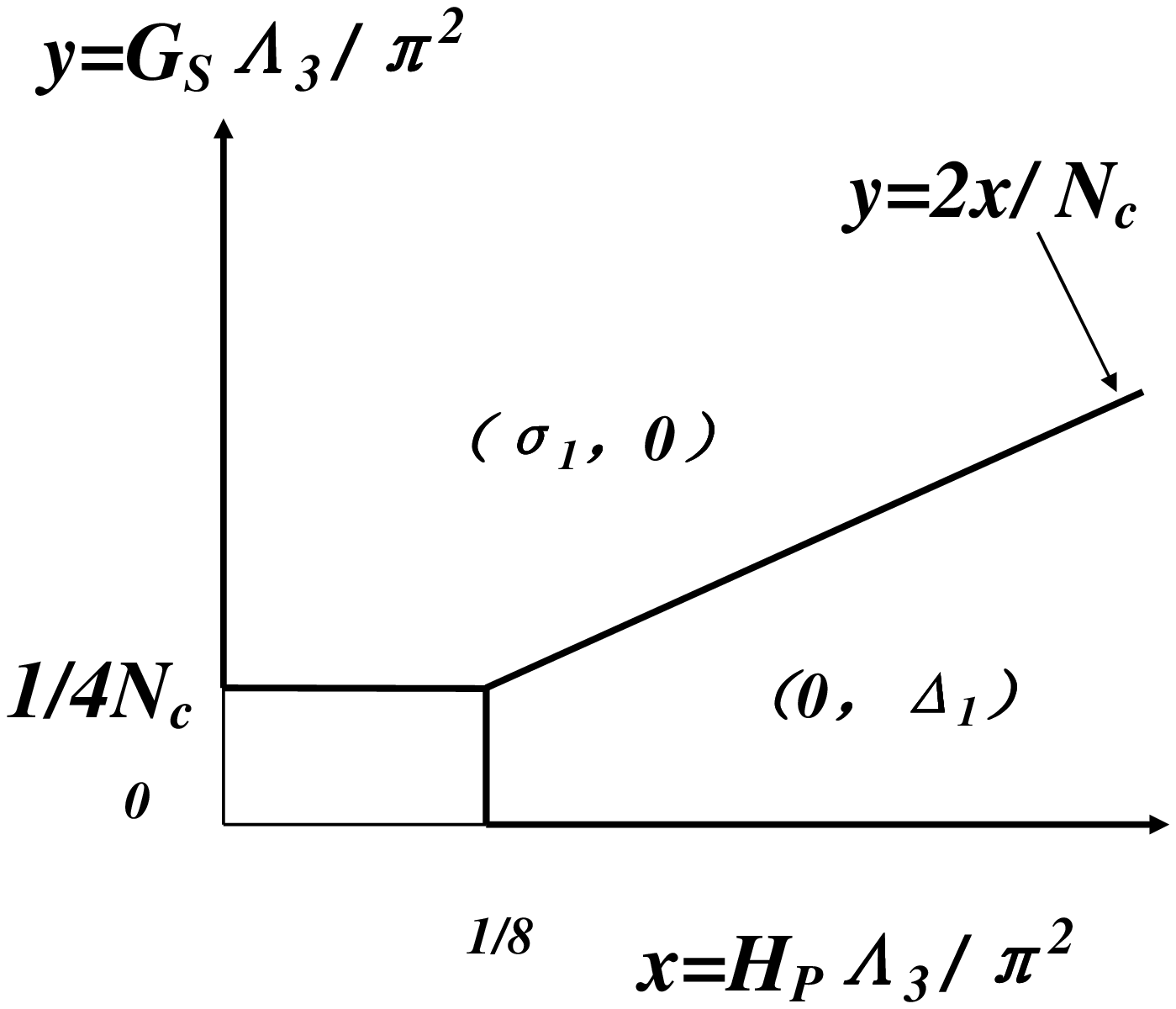}\\
     \vspace{-5cm}{\footnotesize Fig.3: $G_S$-$H_P$ phase diagram of a 3D Gross-Neveu model when $N_c\leq4$ with
     the denotations that $(0,\Delta_1)$--pure $\langle
     qq\rangle$ phase and $(\sigma_1,0)$--pure $\langle
     \bar{q}q\rangle$ phase.
     }}
\end{figure}
\begin{figure}
     {\centering
     \includegraphics[width=0.5\textwidth, height=11cm]{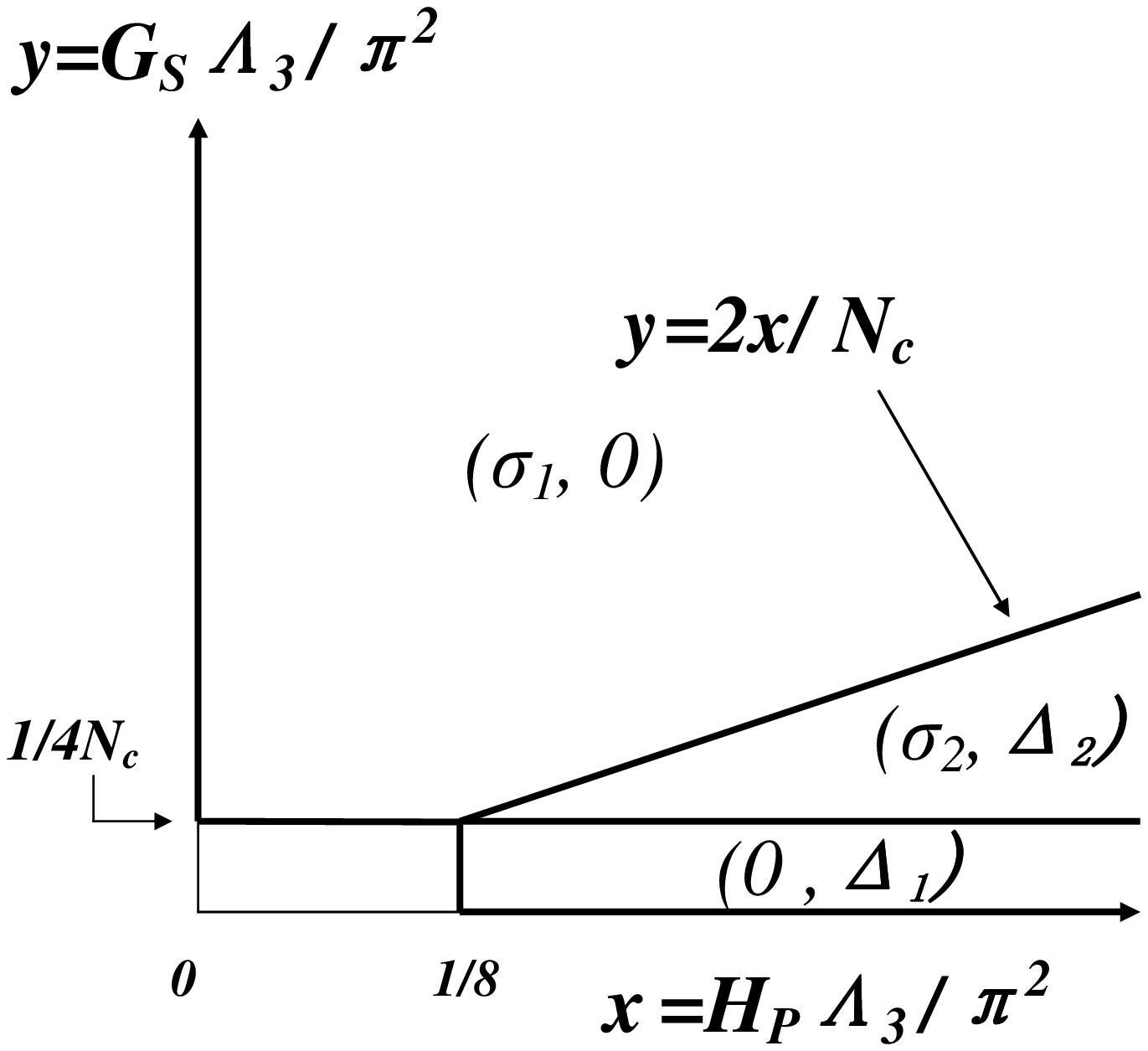}\\
     \vspace{-5cm}{\footnotesize Fig.4: $G_S$-$H_P$ phase diagram of a 3D Gross-Neveu model when $N_c\geq5$
      with the denotations that $(0,\Delta_1)$--pure $\langle
     qq\rangle$ phase, $(\sigma_2,\Delta_2)$--mixed phase with $\langle\bar{q}q\rangle$ and
     $\langle qq\rangle$ and $(\sigma_1,0)$--pure
     $\langle\bar{q}q\rangle$
     phase.
     }}
\end{figure}

 It can be seen from Eq.(34) and Figs.3-4 that,
different from the 4D NJL and 2D GN model, the qualitative feature
of fermionic condensates appearing in the ground state obviously
depends on $N_c$. When the necessary conditions
$G_S\Lambda_3>\pi^2/4N_c$ and $H_P\Lambda_3>\pi^2/8$ are satisfied,
the ground state of the system could be in a pure quark-antiquark
condensate phase only if $G_S/H_P>2/N_c$ for all $N_c$, a similar
result to the one obtained in the 4D and 2D model. However, if
$G_S/H_P<2/N_c$, then whether the coexistence phase with the
condendates $\langle \bar{q}q\rangle$ and $\langle qq\rangle$ could
appear will depend on $N_c$. The coexistence phase could not exist
if $N_c\leq4$ and could do only if $N_c\geq5$ in a 3D GN model. In
addition, also similar to the 4D and 2D model, as $N_c$ increases,
the regions in the phase diagram occupied by the phases with the
diquark condensates including the pure diquark condensate phase and
the coexistence phase with quark-antiquark and diquark condensate
will decrease gradually, and finally go to zeroes as
$N_c\rightarrow\infty$.

\section{Conclusions}\label{conclu}
We have researched interplay between the quark-antiquark and the
diquark condensates in vacuum in two-flavor four-fermion interaction
models with any color number $N_c$. The results are qualitatively
consistent with the ones in $N_c=3$ case in 4D NJL and 2D GN model,
however different from the ones in $N_c=3$ case in 3D GN model. It
has been found that if the ratio of the four-fermion couplings of
$(\bar{q}q)^2$-form and $(qq)^2$-form $G_S/H_S\;
(\mathrm{or}\;G_S/H_P) >2/N_c$, the ratio of the color numbers of
quarks entering the condensates $\langle qq\rangle$ and $\langle
\bar{q}q\rangle$, (and also with sufficiently large $G_S$ in 4D and
3D model), then only the pure $\langle \bar{q}q\rangle$ phase may
exist and this conclusion is $N_c$-independent. Below $2/N_c$, (and
also with sufficiently large $H_S$ or $H_P$ in 4D or 3D model), one
will always first have a mixed phase with the condensates $\langle
\bar{q}q\rangle$ and $\langle qq\rangle$, then a pure $\langle
qq\rangle$ phase, except in the 3D GN model where the mixed phase
does not appear if $N_c\leq4$, the latter includes the known result
in $N_c=3$ case. A common characteristic in all the models is that
as $N_c$ increases, the phases with the diquark condensates $\langle
qq\rangle$ including the pure $\langle qq\rangle$ phase and the
mixed phase with the condensates $\langle \bar{q}q\rangle$ and
$\langle qq\rangle$ will occupy smaller and smaller region in
corresponding phase diagram and finally disappear as
$N_c\rightarrow\infty$. In that limit only the quark-antiquark
condensate phase become possible.\\
\indent The above results certainly deepen our theoretical
understanding of four-fermion interaction models. They not only may
give a definite theoretical constraint on structure of some
phenomenological models \cite{kn:12} but also could provide a useful
clue to the feature of these models at finite temperature and finite
quark chemical potential, as was shown in the 3D GN model with
$N_c=3$.

\end{document}